# Non-Associative Learning Representation in the Nervous System of the Nematode Caenorhabditis elegans

Ramin M. Hasani[*], Magdalena Fuchs[*], Victoria Beneder[**] and Radu Grosu[*]

[*]Cyber-Physical-Systems Group, Vienna University of Technology, Austria
[**]University of Natural Resources and Life Sciences, Vienna, Austria
{ramin.hasani, radu.grosu}@tuwien.ac.at
magdalena.fuchs@student.tuwien.ac.at
victoria.beneder@students.boku.ac.at

March 25, 2017

**Abstract**

*Caenorhabditis elegans* (*C. elegans*) illustrated remarkable behavioral plasticities including complex non-associative and associative learning representations. Understanding the principles of such mechanisms presumably leads to constructive inspirations for the design of efficient learning algorithms. In the present study, we postulate a novel approach on modeling single neurons and synapses to study the mechanisms underlying learning in the *C. elegans* nervous system. In this regard, we construct a precise mathematical model of sensory neurons where we include multi-scale details from genes, ion channels and ion pumps, together with a dynamic model of synapses comprised of neurotransmitters and receptors kinetics. We recapitulate mechanosensory habituation mechanism, a non-associative learning process, in which elements of the neural network tune their parameters as a result of repeated input stimuli. Accordingly, we quantitatively demonstrate the roots of such plasticity in the neuronal and synaptic-level representations. Our findings can potentially give rise to the development of new bio-inspired learning algorithms.

## 1 Learning the Learning Mechanisms

Origin of many deep learning techniques can be traced back to nature [LeCun et al., 2015]. For instance, convolutional neural nets [LeCun et al., 1990], are inspired by the concept of simple and complex cells in the cat's visual cortex [Hubel and Wiesel, 1962]. In this study, we further investigate the sources of learning within the nervous system of small species such as *C. elegans*, by deriving detailed mathematical models of the elements of the nervous system. *C. elegans* is a 1mm-long, bodily transparent worm which dwells in soil and is considered the world's best understood organism [Ardiel and Rankin, 2010]. An adult's stereotypic nervous system comprises exactly 302 neurons connected through 8000 chemical and electrical synapses [White et al., 1986]. The worm, at first, seemed to be a hardwired model organism that can simply crawl forward and backward. Later, it proved otherwise by illustrating fantastic behavioral plasticity. Evidence of expressing well-founded non-associative and associative learning behaviors has been observed in *C. elegans*. Examples include habituation (short term and long term memory) to mechanical stimulation [Wicks and Rankin, 1997, Kindt et al., 2007, Cai et al., 2009], in the context of non-associative learning, and decision making over various favorable environmental stimuli such as food, temperature and oxygen in the form of associative learning [Saeki et al., 2001, Mohri et al., 2005, Tsui and van der Kooy, 2008].

*C. elegans* is highly responsive to experience. This indicates that learning can be mediated from every sensory modality. For instance, mechanosensory habituation is a plasticity for the tap-withdrawal (TW) neural circuit (a neural network responsible for generating a reflexive response as a result of a mechanical tap on the petri dish in which the worm swims [Islam et al., 2016]), where the amplitude and the speed of the reflex can be modified as a result of periodic tap stimulation [Wicks and Rankin, 1997]. Detailed studies on such non-associative type of learning in the TW circuit, reveal fundamental notes underliying learning principles in the *C. elegans*, such as:

- Sensory (input) neurons within the network are subjected to a mediation during the non-associative training process (repeated tap stimulation) [Kindt et al., 2007].

- A single gene's function during learning can interact with multiple postsynaptic connections. Such genes can potentially induce synaptic reconfiguration in several layers during memory consolidation [Stetak et al., 2009].

- Effects of certain genes are local within a neuron; They do not influence the synaptic pathways distributed from that sensory neuron [Ohno et al., 2014].

- During the learning process, change in the amount of neurotransmitter release is notable and is an indicator of learning and memory [Jin et al., 2016].

- Within a neural circuit, only some of the interneurons are proposed to be the substrate of memory [Sugi et al., 2014].

Biological experiments on the principles of learning in *C. elegans*, from high-level behavioral features, to the dynamics of neural circuits, to the gene functional effect on the characteristic of ion channels in a neuron, have demonstrated the remarkable capability of *C. elegans* for learning and memory [Ardiel and Rankin, 2010].

In the present study, we aim to mathematically model the fundamental roots of learning mechanisms within the brain of the worm. Our findings in such attempt presumably enable machine-learning experts to get grounding inspirations towards development of novel learning algorithms. We utilize SIM-CE [Hasani et al., 2017], our simulation platform for performing physiological analysis on neurons and synapses, to explore the non-associative learning principles originated from autonomous gene modifications, variation of neurotransmitters concentrations and mediation of the ion channels activation and deactivation rates. Accordingly, we hypothesize the computational origins of such class of learning within two paradigms of neuronal habituation and synaptic plasticities.

## 2 Neuronal Habituation

In neurons, electrical signals are being generated and propagate as a result of transportation of ions through ion channels [Salkoff et al., 2005]. Figure 1A graphically represents a simplified structure of a neuron of *C. elegans* in which we include voltage-gated calcium channel, calcium pump together with two different potassium channels and a leak channel to be the main mechanism responsible for the dynamics. Neurons can get excited by external sensory stimulus or chemical and electrical synapses from a presynaptic neuron. Considering the total membrane current, one can compose kinetics of the membrane potential and current as follows:

$$C_m \frac{dV}{dt} = -(I_{Ca} + I_K + I_{sk} + I_{Leak}) + \Sigma(I_{Syn} + I_{gap} + I_{stimuli}), \tag{1}$$

where $C_m$ is the membrane capacitance and $V$ represents the membrane potential. $I_{Ca}$, $I_K$, $I_{sk}$ and $I_{Leak}$ stand for the calcium current, potassium current, intercellular calcium-gated potassium channel current and leakage current respectively [Kuramochi and Iwasaki, 2010]. Each ion channel's dynamics is precisely modeled with an inspiration from Hodgkin-Huxley channel modeling technique where the current-voltage relationship is developed based on Ohm's law, $I_{ion} = G_{ion}(E_{ion} - V)$, while conductance of the ion channel, $G_{ion}$ is considered to be the product of a constant maximum conductance with a dynamic parameter which is a function of the membrane potential [Hodgkin and Huxley, 1952].

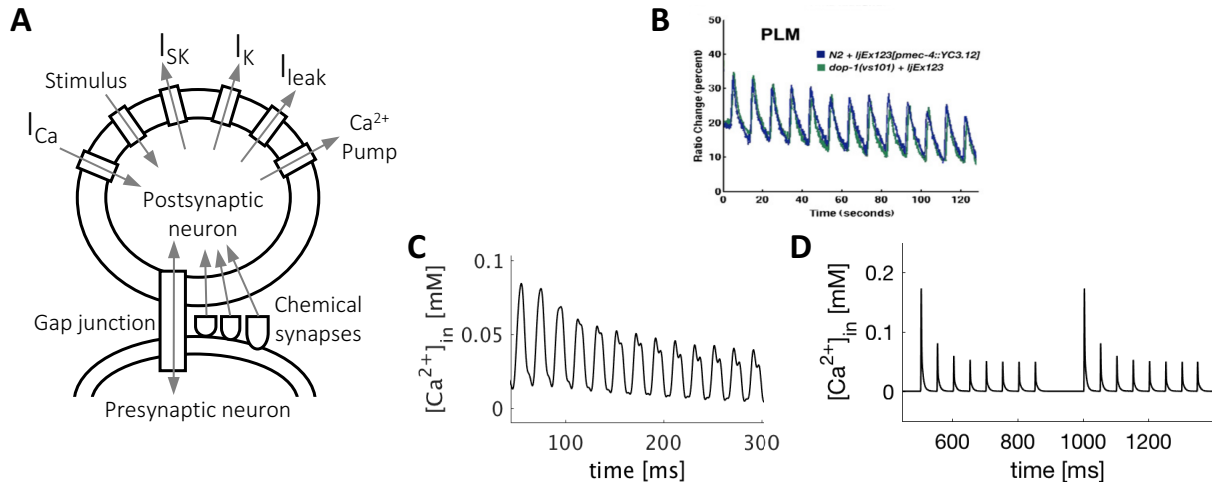

Figure 1: Learning the learning representation. A) Structure of a neuron B) Habituation of a real sensory neuron (reprinted with permission from [Kindt et al., 2007]) C) Model of a neuronal habitation D) Response of an interneuron habituated and dishabituated with synaptic plasticities.

In order to model the gene modification effects and the corresponding learning, we consider the maximum conductance of some ion channels to be a dynamic variable as a function of the input stimulus maximum value in time. In order to evaluate our assumption, we compare our result to the real sensory neuron habituation. A real sensory neuron habituates to repeated mechanical pulses on the body of the worm. The effect appears in the form of a reduction in the level of intracellular calcium concentration shown in Figure 1B. In our simulation such behavior is captured by modifying the maximum potassium channel conductance to be a dynamic function of the input tap-stimulus and time (Figure 1C).

**Key finding 1**: The state of the activation of an input neuron can be modified by repeatedly induced stimuli through an intrinsic parameter which is dynamically altered based on the input amplitude and frequency.

## 2.1 Synaptic Plasticity

Fundamentally, many learning mechanisms originate from the alternation of the amount of neurotransmitters concentration. Information propagates from a postsynaptic neuron to a presynaptic cell through chemical and electrical synapses. The information transportation process depends on several circumstances; the amount of available neurotransmitters, $G_{max}$, at the presynaptic end, the state of the activation of the presynaptic cell which causes the neurotransmitter secretion tone $G(V_{pre})$, number of available receptors at the postsynaptic neuron and the probability of the neurotransmitters binding to the postsynaptic receptors, $S(t)$. A detailed model of a chemical synapse's current therefore can be derived as follows:

$$I_{Syn} = G_{max}G(V_{pre})S(t)(E_{Syn} - V_{post}), \qquad (2)$$

where $E_{Syn}$ represents the reversal potential of a synapse [Schutter, 2009]. Variables in Eq. 2 are dynamically modeled by partial differential equations where the steady state and the time-constant of each variable can be tuned in order to capture synaptic plasticity in *C. elegans*. Rankin and Wicks [2000] showed that the glutamate concentration considereably modifies the habituation responses in touch sensory neurons. Such behavior is captured as a property of a synapse where simultaneous alternation of the parameters inside $G_{max}$ and $S(t)$, results in the similar tap-withdrawal response (See Figure 1D).

**Key finding 2**: Synapses with an internal state can autonomously modify the behavior of the system. Synapses can presumably be involved in completion of a habituation process, dishabituation process (See Figure 1D the second calcium spike train) and propagation of a neuronal habituation to the rest of the neural circuit.

## 3 Final Note

Learning is an essential attribute for survival. *C. elegans* demonstrates optimal adaptive behavior to react to the environmental circumstances for survival. Learning how the animal learns and adapts not only is a notable step forward towards decoding the human brain's principles but can also lead us to the development of better learning algorithms.[1]

---

[1] A comprehensive version of this work, will be published soon elsewhere.

# Model of Learning in C. elegans